\documentstyle[preprint,eqsecnum,aps]{revtex}
\tightenlines
\newcommand{\be}{\begin{equation}}
\newcommand{\ee}{\end{equation}}
\newcommand{\bea}{\begin{eqnarray}}
\newcommand{\eea}{\end{eqnarray}}
\newcommand{\ba}{\begin{array}}
\newcommand{\ea}{\end{array}}

\newcommand{\th}{\theta}

\newcommand{\ga}{\gamma}

\newcommand{\la}{\lambda}

\newcommand{\Om}{\Omega}

\newcommand{\pa}{\partial}

\newcommand{\no}{\nonumber}

\newcommand{\res}{\mbox{res}}

\begin{document}
\draft
\title{On Darboux-B\"acklund Transformations for the \\
Q-Deformed Korteweg-de Vries Hierarchy}
\author{Ming-Hsien Tu$^1$, Jiin-Chang Shaw$^2$  and Chin-Rong Lee$^3$}

\address{
$^1$ Department of Mathematics, 
National Chung Cheng University, \\
Minhsiung, Chiayi 621, Taiwan\\
$^2$ Department of Applied Mathematics, 
National Chiao Tung University, \\
Hsinchu 300, Taiwan\\
$^3$ Department of Physics, 
National Chung Cheng University, \\
Minhsiung, Chiayi 621, Taiwan}
\date{\today}
\maketitle
\begin{abstract}
We study Darboux-B\"acklund  transformations (DBTs)
for the $q$-deformed Korteweg-de Vries hierarchy by
using the $q$-deformed pseudodifferential operators. 
We identify the elementary DBTs which are triggered by the 
gauge operators constructed from
the (adjoint) wave functions of the hierarchy. Iterating these
elementary DBTs we obtain not only $q$-deformed
Wronskian-type but also binary-type representations of the 
tau-function to the hierarchy.
\end{abstract}
\newpage
\section{Introduction}

There has been remarkable interest in the theory of soliton
ever since the discovery of the Inverse Scattering Method
(ISM) for the Korteweg-de Vires(KdV) equation(see, for example, \cite{NMPZ}). 
Although the ISM has been extended to many nonlinear systems 
which describe phenomena in many branches of science, 
the KdV equation still plays an important role in the development 
of modern soliton theory (see, for example \cite{95}).
In particular, many concepts were established first for the KdV equation 
and then generalized to other systems in different ways. 

A convenient approach to formulate the KdV hierarchy relies on 
the use of fractional-power pseudo-differential operators associated 
with the scalar Lax operator $L=\pa^2+u$ which was further generalized 
by Gelfand and Dickey \cite{D} to the $N$-th KdV hierarchy which has Lax 
operator of the form $L_N=\pa^N+u_{N-1}\pa^{N-1}+\cdots+u_0$.
In the past few years, there are several works concerning the extensions of 
the KdV hierarchy in Lax formulation, such as Drinfeld-Skolov theory \cite{DS}
and supersymmetric generalizations \cite{MR}, etc.
The common features of these extensions show that they preserve 
the integrable structure of the KdV hierarchy and contain the KdV 
as reductions in some limiting cases .

Recently, a new kind of extension called $q$-deformed KdV ($q$-KdV)
hierarchy has been proposed and studied \cite{Z,JM,F,FR,KLR,HI,Ia,Ib,AHM}. 
In this extension the partial
derivative $\pa$  is replaced by the $q$-deformed differential operators ($q$-DO)
$\pa_q$ such that
\be
(\pa_qf(x))=\frac{f(qx)-f(x)}{x(q-1)}
\label{diff}
\ee
which recovers the ordinary differentiation $(\pa_xf(x))$ as $q$ goes to 1.
Even though many structures of the $q$-KdV hierarchy 
such as infinite conservation laws \cite{Z},
bi-Hamiltonian structure \cite{JM}, tau-function \cite{HI,Ia,Ib,AHM}, 
Vertex operators\cite{AHM}, Virasoro and $W$-algebras \cite{FR}, etc.
have been studied however the $q$-version of Darboux-B\"acklund 
transformations (DBTs) for this system are still unexplored. 
It is well known that the DBT is an important property to characterize
the integrability of the hierarchy\cite{MS}. 
Thus it is worthwhile to investigate the DBTs associated with 
the $q$-KdV hierarchy. Once this goal can be achieved, it will deepen our 
understanding on the soliton solutions of the hierarchy.

Our paper is organized as follows: In Sec. II, we recall 
the basic facts concerning the $q$-deformed pseudodifferential 
operators ($q$-PDO)
and define the $N$th $q$-deformed Korteweg-de Vries 
(q-KdV) hierarchy. In Sec. III, we construct the 
Darboux-B\"acklund transformations (DBTs) for the $N$th
q-KdV hierarchy, which preserve the form of the Lax operator 
and the hierarchy flows. Iteration of these DBTs generates 
the $q$-analogue of soliton solutions of the hierarchy. 
In Sec. IV, the case for $N=2$ ($q$-KdV hierarchy) is studied
in detail to illustrate the $q$-deformed formulation. 
Concluding remarks are presented in Sec. V.

\section{$q$-deformed pseudodifferential operators}

Let us define the $q$-shift operator $\th$ as
\be
\th(f(x))=f(qx)
\label{shift}
\ee
then it is easy to show that $\th$ and $\pa_q$ do not commute 
but satisfy
\be
(\pa_q\th^k(f))=q^k\th^k(\pa_qf),\qquad k\in Z
\ee
Using (\ref{diff}) and (\ref{shift})  we have 
$(\pa_qfg)=(\pa_qf)g+\th(f)(\pa_qg)$
which implies that
\be
\pa_qf=(\pa_qf)+\th(f)\pa_q
\ee
We also denote $\pa_q^{-1}$ as the formal inverse of $\pa_q$ 
such that $\pa_q\pa_q^{-1}f=\pa_q^{-1}\pa_qf=f$ and hence
\be
\pa_q^{-1}f=\sum_{k\geq 0}(-1)^kq^{-k(k+1)/2}\th^{-k-1}(\pa_q^kf)
\pa_q^{-k-1}.
\ee
In general one can justify the following $q$-deformed Leibnitz 
rule:
\be
\pa_q^nf=\sum_{k\geq 0}{n \brack k}_q\th^{n-k}(\pa_q^kf)\pa_q^{n-k}\qquad
n\in Z
\ee
where we introduce the $q$-number and the $q$-binomial as follows
\bea
[n]_q&=&\frac{q^n-1}{q-1}\\
{n \brack k}_q&=&\frac{[n]_q[n-1]_q\cdots[n-k+1]_q}
{[1]_q[2]_q\cdots[k]_q},\qquad {n \brack 0}_q= 1
\eea
For a $q$-PDO of the form
\be
P=\sum_{i=-\infty}^nu_i\pa_q^i
\ee
it is convenient to separate $P$ into the differential 
and integral parts as follows
\be
P_+=\sum_{i\geq 0}u_i\pa_q^i\qquad P_-=\sum_{i\leq -1}u_i\pa_q^i
\ee
and denote $(P)_0$ as the zero-th order term of $P$.
The residue of $P$ is defined by $\res(P)=u_{-1}$ and the conjugate 
operation ``$*$" for $P$ is defined by $P^*=\sum_i(\pa_q^*)^iu_i$
with
\be
\pa_q^*=-\pa_q\th^{-1}
\ee
Then a straightforward calculation shows that 
\be
(PQ)^*=Q^*P^*
\ee
where $P$ and $Q$ are any two $q$-PDOs.
Finally, for a set of functions $f_1,f_2,\cdots,f_n$,
we define $q$-deformed Wronskian determinant
$W_q[f_1,f_2,\cdots,f_n]$ as
\be
W_q[f_1,f_2,\cdots,f_n]=
\left|
\ba{ccc}
f_1 & \cdots & f_n\\
(\pa_qf_1)& \cdots &(\pa_qf_n)\\
\vdots& &\vdots\\
(\pa_q^{n-1}f_1)& \cdots &(\pa_q^{n-1}f_n)
\ea
\right|
\ee

With these definitions in hand, we have the following 
identities which will simplify the computations involving
compositions of $q$-PDOs.

{\bf Proposition 1:\/} 
\bea
& &(P^*)_+=(P_+)^*,\qquad (P^*)_-=(P_-)^* \\
& &(\pa_q^{-1}P)_-=\pa_q^{-1}(P^*)_0+\pa_q^{-1}P_-\\
& &(P \pa_q^{-1})_-=(P)_0\pa_q^{-1}+P_-\pa_q^{-1}\\
& &\res(P^*)=-\th^{-1}(\res(P)),\qquad (\pa_q\res(P))=\res(\pa_qP)
-\th(\res(P \pa_q))\\
& &\res(P \pa_q^{-1})=(P)_0\qquad \res(\pa_q^{-1}P)=\th^{-1}((P^*)_0)\\
& &\res(\pa_q^{-1}P_1P_2\pa_q^{-1})
=\res(\pa_q^{-1}(P_1^*)_0P_2\pa_q^{-1})+\res(\pa_q^{-1}P_1(P_2)_0\pa_q^{-1})
\eea
where $P_1=(P_1)_+$ and $P_2=(P_2)_+$.

The $N$-th $q$-KdV hierarchy is defined, in Lax form, as
\be 
\pa_{t_n}L=[B_n, L],\qquad n=1,2,3,\cdots
\label{laxeq}
\ee
with
\be
L=\pa_q^N+u_{N-1}\pa_q^{N-1}+\cdots+u_0,
\qquad  B_n\equiv L^{n/N}_+
\label{lax}
\ee
where the coefficients   $u_i$  are functions of the variables $(x,t_1,t_2,\cdots)$
but do not depend on $(t_N,t_{2N},t_{3N},\cdots)$.

In fact, we can rewrite the hierarchy equations (\ref{laxeq}) as follows:
\be
\pa_{t_m}B_n-\pa_{t_n}B_m-[B_n,B_m]=0
\label{zero}
\ee
which is called the zero-curvature condition and is equivalent to
the whole set of equations of (\ref{laxeq}). 
If we can find a set of functions $\{u_i, i=0,1,\cdots,N-1\}$ and 
hence a corresponding Lax operator $L$ (or $B_n$)
satisfying (\ref{laxeq}) (or (\ref{zero})), 
then we have a solution to the $N$-th $q$-KdV hierarchy.

For the Lax operator (\ref{lax}), we can formally expand $L^{1/N}$ 
in powers of $\pa_q$ as follows
\be
L^{1/N}=\pa_q+s_0+s_1\pa_q^{-1}+\cdots
\ee
such that $(L^{1/N})^N=L$ which gives all the $s_i$ being 
$q$-deformed differential polynomials in $\{u_i\}$. 
Especially, for the coefficient of $\pa_q^{N-1}$ we have
\be
u_{N-1}=s_0+\th(s_0)+\cdots+\th^{N-1}(s_0).
\ee

The Lax equations (\ref{laxeq}) can be viewed as the 
compatibility condition of the linear system
\be
L\phi=\la\phi,\qquad \pa_{t_n}\phi=(B_n\phi)
\label{wave}
\ee
where $\phi$ and $\la$ are called wave function and eigenvalue of 
the linear system, respectively. 
On the other hand, we can also introduce
adjoint wave function $\psi$ which satisfies the adjoint
linear system
\be
L^*\psi=\mu\psi,\qquad \pa_{t_n}\psi=-(B_n^*\psi)
\label{adwave}
\ee
For convenience, throughout this paper, $\phi_i$ $(\psi_i)$ 
will stand for (adjoint) wave functions with eigenvalues $\la_i (\mu_i)$, 
respectively without further mention.

\section{elementary Darboux-B\"acklund transformations}

In this section we would like to construct DBTs for 
the $N$-th $q$-KdV hierarchy. 
To attain this purpose, let us consider the following transformation
\be
L\to L^{(1)}=TLT^{-1}
\label{laxgt}
\ee
where $T$ is any reasonable $q$-PDO and $T^{-1}$
denotes its inverse. 
In order to obtain the new solution ($L^{(1)}$) from the old 
one ($L$) the gauge operator $T$ can not be arbitrarily chosen.
It should be constructed in such a way that the transformed Lax 
operator $L^{(1)}$ preserves the form of $L$ and satisfies 
the Lax equation (\ref{laxeq}). 
From the zero-curvature condition (\ref{zero})
point of view, the operator $B_n$ should be transformed according to
\be
B_n\to B^{(1)}_n=TB_nT^{-1}+\pa_{t_n}TT^{-1}
\label{bngt}
\ee
which will, in general, not be a pure $q$-DO
although the $B_n$ does. However if we suitable choose the gauge 
operator $T$ such that $B^{(1)}_n$, as defined by (\ref{bngt}), is also a 
purely $q$-PDO, then $L_n^{(1)}$ 
represents a valid  new solution to the $N$-th $q$-KdV hierarchy. 
This is the goal we want to achieve in this letter.

To formulate the DBTs, we follow \cite{O} to  introduce a $q$-version 
of the bilinear potential $\Om(\phi,\psi)$ which is constructed from a wave 
function $\phi$ and an adjoint wave function $\psi$.
The usefulness of this bilinear potential will
be clear when we use it to construct DBTs (see below).

{\bf Proposition 2:\/} For any pair of $\phi$ and $\psi$, 
there exists a bilinear potential $\Om(\phi, \psi)$ 
satisfying
\bea
(\pa_q\Om(\phi, \psi))&=&\phi\psi\\
\pa_{t_n}\Om(\phi, \psi)&=&\res(\pa_q^{-1}\psi B_n\phi \pa_q^{-1})
\eea
In fact the bilinear potential $\Om(\phi, \psi)$ can be formally
represented  by a $q$-integration as $\Om(\phi,\psi)=(\pa_q^{-1}\phi\psi)$.

Motivated by the DBTs for the ordinary KdV \cite{MS} 
(or Kadomtsev-Petviashvili (KP) \cite{O,OS,CSY}) 
hierarchy, we can construct a qualified gauge operator $T$ as follows
\be
T_1=\th(\phi_1)\pa_q\phi_1^{-1}=\pa_q-\alpha_1,\qquad 
\alpha_1\equiv \frac{(\pa_q\phi_1)}{\phi_1}
\label{tdb}
\ee
where $\phi_1$ is a wave function associated with the linear 
system (\ref{wave}). It is not hard to show that the transformed 
Lax operator $L^{(1)}$ is a purely $q$-PDO with order 
$N$ and the Lax equation (\ref{laxeq}) transforms covariantly, i.e. 
$\pa_{t_n}L^{(1)}=[(L^{(1)})_+^{n/N}, L^{(1)}]$. 
The transformed coefficients $\{u_i^{(1)}\}$
then can be expressed in terms of $\{u_i\}$ and $\phi_1$. 
On the other hand, for a given generic wave function $\phi\neq \phi_1$ 
(or adjoint function $\psi$) its transformed result can be expressed 
in terms of $\phi_1$ and itself.

{\bf Proposition 3:\/} Suppose $\phi_1$ is a wave function of the 
linear system (\ref{wave}). 
Then the gauge operator $T_1$ triggers the following DBT:
\bea
& &L^{(1)}=T_1LT_1^{-1}=
\pa_q^N+u_{N-1}^{(1)}\pa_q^{N-1}+\cdots+u_0^{(1)}\\
& &\phi^{(1)}=(T_1\phi)=\frac{W_q[\phi_1, \phi]}{\phi_1},\qquad
\phi\neq \phi_1\\
& &\psi^{(1)}=((T_1^{-1})^*\psi)=-\frac{\th(\Om(\phi_1, \psi))}{\th(\phi_1)}
\eea
where $L^{(1)}$, $\phi^{(1)}$ and $\psi^{(1)}$ 
satisfy (\ref{laxeq}), (\ref{wave}) and (\ref{adwave}) respectively. 

Just like the DBTs for the ordinary KdV hierarchy, the DBT triggered  by 
the gauge operator $T_1$ is by no means the only transformation in this 
$q$-analogue framework. We have another construction  of DBT by 
using the adjoint wave function associated with the adjoint linear 
system (\ref{adwave}). Indeed, for a given adjoint wave 
function $\psi_1$ we can construct a gauge operator 
\be
S_1=\th^{-1}(\psi_1^{-1})\pa_q^{-1}\psi_1=(\pa_q+\beta_1)^{-1},
\qquad \beta_1\equiv \frac{(\pa_q\th^{-1}(\psi_1))}{\psi_1}
\label{sdb}
\ee
which preserves the form of the Lax operator and the Lax equation.

{\bf Proposition 4:\/} Suppose $\psi_1$ is an adjoint wave function of the 
adjoint linear system (\ref{adwave}). 
Then the gauge operator $S_1$ triggers the following adjoint DBT:
\bea
& &L^{(1)}=S_1LS_1^{-1}\\
& &\phi^{(1)}=(S_1\phi)=\frac{\Om(\phi, \psi_1)}{\th^{-1}(\psi_1)}\\
& &\psi^{(1)}=((S_1^{-1})^*\psi)=\frac{\tilde{W}_q[\psi_1, \psi]}{\psi_1},
\qquad \psi\neq \psi_1
\eea
where $L^{(1)}$, $\phi^{(1)}$ and $\psi^{(1)}$ 
satisfy (\ref{laxeq}), (\ref{wave}) and (\ref{adwave}), respectively and 
$\tilde{W}_q$ is obtained from $W_q$ by replacing $\pa_q$ with $\pa_q^*$.

So far, we have constructed two elementary DBTs triggered by the
gauge operators $T_1$ and $S_1$. Regarding them as the building
blocks, we can construct more complicated transformations from
the compositions of these elementary DBTs. However, we will see
that it is convenient to consider a DBT followed by an adjoint
DBT and vice versa because such combination will frequently
appear in more complicated DBTs. So let us compose
them to form a single operator $R_1$ which we call binary gauge operator.
The construction of the binary gauge operator $R_1$ can be realized as follows: 
first we perform a DBT triggered by the gauge operator 
$T_1=\th(\phi_1)\pa_q\phi_1^{-1}$ and the adjoint
wave function $\psi_1$ is thus transformed to 
$\psi_1^{(1)}=((T_1^{-1})^*\psi_1)=-\th(\phi_1^{-1})\th(\Om(\phi_1, \psi_1))$. 
Then a subsequent adjoint DBT  triggered by 
$S_1^{(1)}=\th^{-1}((\psi_1^{(1)})^{-1})\pa_q^{-1}(\psi_1^{(1)})$ 
is performed and the composition of these two transformations gives 
\be
R_1=S_1^{(1)}T_1=1-\phi_1\Om(\phi_1,\psi_1)^{-1}\pa_q^{-1}\psi_1
\label{rdb}
\ee

{\bf Proposition 5:\/} Let $\phi_1$ and $\psi_1$ be wave function and 
adjoint wave function associated with the linear 
systems (\ref{wave}) and (\ref{adwave}), respectively. 
Then the gauge operator $R_1$ triggers the following binary DBT:
\bea
& &L^{(1)}=R_1LR_1^{-1}\\
& &\phi^{(1)}=(R\phi)=\phi-\Om(\phi_1,\psi_1)^{-1}\phi_1\Om(\phi,\psi_1),
\qquad \phi\neq \phi_1\\
& &\psi^{(1)}=((R^{-1})^*\psi)=
\psi-\th(\Om(\phi_1,\psi_1)^{-1})\psi_1\th(\Om(\phi_1,\psi)),
\qquad \psi\neq \psi_1
\eea
where $L^{(1)}$, $\phi^{(1)}$ and $\psi^{(1)}$ 
satisfy (\ref{laxeq}), (\ref{wave}) and (\ref{adwave}) respectively.

We would like to remark that the construction of the binary gauge 
operator $R_1$ is independent of the order of the gauge 
operators $T$ and $S$. If we apply  the gauge operator $S_1$ followed 
by $T_1^{(1)}$, then a direct calculation shows that $R=T_1^{(1)}S_1$ 
has the same form as (\ref{rdb}).

The remaining part of this section is to consider the iteration of the DBTs 
by using the DBT, the adjoint DBT, and the binary DBT 
triggered by the gauge operators $T$, $S$, and $R$, respectively.
For example, by iterating the DBT triggered by the gauge operator $T$,
we can express the solution of the $N$-th $q$-KdV hierarchy through
the $q$-deformed Wronskian representation. This construction
starts with $n$ wave functions $\phi_1, \phi_2,\cdots, \phi_n$
of the linear system (\ref{wave}). Using $\phi_1$, say, to perform
the first DBT of Proposition 1, then all $\phi_i$ are transformed
to $\phi_i^{(1)}=(T_1\phi_i)$. Obviously, we have $\phi_1^{(1)}=0$.
The next step is to perform a subsequent DBT triggered by $\phi_2^{(1)}$,
which leads to the new wave functions $\phi_i^{(2)}$ with $\phi_2^{(2)}=0$.
Iterating this process such that all the wave functions are exhausted,
then an $n$-step DBT triggered by the gauge operator 
$T_n=(\pa_q-\alpha_n^{(n-1)})(\pa_q-\alpha_{n-1}^{(n-2)})
\cdots(\pa_q-\alpha_1)$ is obtained, where 
$\alpha_i^{(j)}\equiv (\pa_q\phi_i^{(j)})/\phi_i^{(j)}$. 
It is easy to see that $T_n$ is an $n$th-order $q$-DO
of the form $T_n=\pa_q^n+a_{n-1}\pa_q^{n-1}+\cdots+a_0$  with  
$a_i$ defined by the conditions 
$(T_n\phi_j)=0, j=1,2,\cdots,n$. Following the Cramer's formula
it turns out that $a_i=-W_q^{(i)}[\phi_1,\cdots,\phi_n]/
W_q[\phi_1,\cdots,\phi_n]$ where $W_q^{(i)}$ is obtained from
$W_q$ with its $i$-th row replaced by 
$(\pa_q^n\phi_1),\cdots,(\pa_q^n\phi_n)$.
This implies that the $n$-step transformed wave function
$\phi^{(n)}$ ($\phi\neq \phi_i$) is given by
\be
\phi^{(n)}=(T_n\phi)=\frac{W_q[\phi_1,\cdots,\phi_n,\phi]}
{W_q[\phi_1,\cdots,\phi_n]}
\label{nwf}
\ee
and the $n$-step gauge operator $T_n$ can be expressed as
\be
T_n=\frac{1}{W_q[\phi_1,\cdots,\phi_n]}
\left|
\ba{cccc}
\phi_1 &\cdots&\phi_n &1 \\
(\pa_q\phi_1) &\cdots&(\pa_q\phi_n) &\pa_q \\
\vdots &  & \vdots& \vdots \\
(\pa_q^n\phi_1) &\cdots&(\pa_q^n\phi_n) &\pa_q^n
\ea
\right|
\ee
where it should be realized that in the expansion of the
determinant by the elements of the last column,
$\pa_q^i$ have to be written to the right of the minors.

Next let us turn to the iteration of the adjoint DBT. In this case,
the $n$-step gauge operator can be constructed in a similar manner
by preparing $n$ initial adjoint wave functions 
$\psi_1,\cdots,\psi_n$ such that
$S^{-1}_n=(\pa_q+\beta_n^{(n-1)})(\pa_q+\beta_{n-1}^{(n-2)})\cdots
(\pa_q+\beta_1)=\pa_q^n+\sum_{i=1}^{n-1}\pa_q^ib_i$.
Using the required conditions $((S^{-1}_n)^*\psi_i)=0, j=1,\cdots,n$
we obtain $b_i=-\tilde{W}_q^{(i)}[\psi_1,\cdots,\psi_n]/
\tilde{W}_q[\psi_1,\cdots,\psi_n]$ 
which give the $n$-step transformed adjoint wave function
\be
\psi^{(n)}=((S_n^{-1})^*\psi)=\frac{\tilde{W}_q[\psi_1,\cdots,\psi_n,\psi]}
{\tilde{W}_q[\psi_1,\cdots,\psi_n]}
\label{nawf}
\ee
and the gauge operator
\be
(S_n^{-1})^*=\frac{1}{\tilde{W}_q[\psi_1,\cdots,\psi_n]}
\left|
\ba{cccc}
\psi_1 &\cdots&\psi_n &1\\
 (\pa_q^*\psi_1) &\cdots&(\pa_q^*\psi_n) &\pa_q^* \\
 \vdots &  & \vdots & \vdots \\
 ((\pa_q^*)^n\psi_1) &\cdots&((\pa_q^*)^n\psi_n)&(\pa_q^*)^n 
\ea
\right|
\ee
Finally, we shall construct $n$-step binary DBT by preparing 
$n$ wave functions $\phi_i$ and $n$ adjoint wave functions $\psi_i$ at the begining.
Then we perform the first binary DBT by using the gauge operator
$R_1$ which is constructed from $\phi_1$ and $\psi_1$ as in 
(\ref{rdb}). At the same time, all $\phi_i$ and $\psi_i$
are transformed to $\phi_i^{(1)}=(R_1\phi_i)$ 
and $\psi_i^{(1)}=((R_1^{-1})^*\psi_i)$, respectively
except $\phi_1^{(1)}=\psi_1^{(1)}=0$. We then use the pair 
$\{\phi_2^{(1)},\psi_2^{(1)}\}$ to perform the next binary DBT.
Iterating this process until all the pairs $\{\phi_i,\psi_i\}$
are exhausted, then we obtain an $n$-step gauge operator of
the form $R_n=1-\sum_{j=1}^{n-1}c_j\pa_q^{-1}\psi_j$.
 Solving the  conditions $(R_n\phi_j)=0, j=1,\cdots,n$,
we obtain the coefficients $c_j=\det\Om^{(i)}/\det \Om$
where $\Om_{ij}\equiv \Om(\phi_i,\psi_j)$ and $\Om^{(i)}$
is obtained from $\Om$ with its $i$-th column replaced by
$(\phi_i,\cdots,\phi_n)^t$. This leads to the following 
representations for $R_n$:
\be
R_n=\frac{1}{\det \Om}
\left|
\ba{cccc}
\Om_{11}&\cdots&\Om_{1n}&\phi_1 \\
\vdots& &\vdots& \vdots \\
\Om_{n1}&\cdots&\Om_{nn}&\phi_n \\
 \pa_q^{-1}\psi_1&\cdots&\pa_q^{-1}\psi_n&1
\ea
\right|
\ee
Moreover, the $n$-step transformed bilinear potential
$\Om(\phi^{(n)},\psi^{(n)})$ can be expressed in terms of 
binary-type determinant as
\be
\Om(\phi^{(n)}, \psi^{(n)})=
\frac{1}{\det \Om}
\left|
\ba{cccc}
\Om_{11} & \cdots&\Om_{1n}&\Om(\phi_1,\psi)\\
\vdots  & &\vdots &\vdots\\
\Om_{n1} & \cdots&\Om_{nn}&\Om(\phi_n,\psi)\\
\Om(\phi, \psi_1) & \cdots&\Om(\phi, \psi_n)&\Om(\phi,\psi)\\
\ea
\right|
\label{nbf}
\ee
where $\phi^{(n)}=(R_n\phi)$ and $\psi^{(n)}=((R_n^{-1})^*\psi)$.

\section{$q$-deformed KdV hierarchy ($N=2$)}

This section is devoted to illustrate the DBTs for the simplest
nontrivial example: $q$-deformed KdV hierarchy ($N=2$ case).
Let 
\be
L=\pa_q^2+u_1\pa_q+u_0
\label{laxkdv}
\ee
then the Lax equations
\be
\pa_{t_n}L=[L^{n/2}_+, L],\qquad n=1,3,5,\cdots
\label{laxeqkdv}
\ee
define the evolution equations for $u_1$ and $u_0$. In particular,
for the $t_1$-flow, we have
\bea
\pa_{t_1}u_1&=&x(q-1)\pa_{t_1}u_0\\
\pa_{t_1}u_0&=&(\pa_qu_0)-(\pa_q^2s_0)-(\pa_qs_0^2)
\eea
which is nontrivial and recovers the ordinary case as $q$ goes to 1.
For  higher hierarchy flows the evolution equations for $u_1$ and $u_0$ 
become more complicated due to the non-commutative nature of 
the $q$-deformed formulation.

We now perform the DBT of Proposition 3 to the Lax operator 
(\ref{laxkdv}), then the transformed coefficients become
\bea
u_1^{(1)}&=&\th(u_1)-\alpha_1+\th^2(\alpha_1)
\label{u1}\\
u_0^{(1)}&=&\th(u_0)+(\pa_qu_1)+(q+1)\th(\pa_q\alpha_1)-
\alpha_1u_1+\th(\alpha_1)u_1^{(1)}
\label{u0}
\eea
Since $\phi_1$ is a wave function associated with the 
Lax operator (\ref{laxkdv}), i.e. $L\phi_1=\la_1\phi_1$,
one can easily verify that
$(\pa_q\alpha_1)+\th(\alpha_1)\alpha_1+u_1\alpha_1+u_0=\la_1$
and hence Eqs.(\ref{u1}) and (\ref{u0}) can be simplified as
\bea
u_1^{(1)}-u_1&=&x(q-1)(u_0^{(1)}-u_0)
\label{u11}\\
u_0^{(1)}-u_0&=&\pa_q(u_1+\alpha_1+\th(\alpha_1))
\label{u01}
\eea
Furthermore, using the facts that 
$\pa_{t_1}\phi_1=(L_+^{1/2}\phi_1)=(\pa_q\phi_1)+s_0\phi_1$ 
and $u_1=\th(s_0)+s_0$, we can rewrite (\ref{u01}) as
\be
u_0^{(1)}=u_0+\pa_q\pa_{t_1}\ln\phi_1\th(\phi_1)
\label{u02}
\ee
Eqs.(\ref{u11}) and (\ref{u02}) are just the desired 
result announced in Sec. III.

Similarly, by applying the above analysis to the
adjoint and binary DBTs, we obtain the following result:

{\bf Proposition 6:\/} Let $\phi_1$ and $\psi_1$ be (adjoint) 
wave function associated with the Lax operator (\ref{laxkdv}).
Then under the DBT, adjoint DBT, and binary DBT, the transformed
coefficients are given by
\be
u_1^{(1)}-u_1=x(q-1)(u_0^{(1)}-u_0)
\ee
with
\bea
u_0^{(1)}&=&u_0+\pa_q\pa_{t_1}\ln\phi_1\th(\phi_1),
\qquad (\mbox{DBT})
\label{1db}\\
u_0^{(1)}&=&u_0+\pa_q\pa_{t_1}\ln\psi_1\th^{-1}(\psi_1)
,\qquad (\mbox{adjoint DBT})
\label{1adb}\\
u_0^{(1)}&=&u_0+\pa_q\pa_{t_1}\ln\Om_{11}\th(\Om_{11})
,\qquad (\mbox{binary DBT})
\label{1bdb}
\eea
Eqs.(\ref{1db})-(\ref{1bdb}) effectively represent the 1-step
transformations. To obtain the $n$-step DBT, adjoint DBT and
binary DBT we just need to  iterate the corresponding 1-step transformations 
successively by inserting the triggered wave function (\ref{nwf}), adjoint wave 
function (\ref{nawf}) and bilinear potential (\ref{nbf}) into the logarithm in Eqs.
(\ref{1db}), (\ref{1adb}) and (\ref{1bdb}), respectively.

{\bf Proposition 7:\/} Let $\phi_i$ and $\psi_i$ 
($i=1,2,\cdots,n$) be (adjoint) 
wave functions associated with the Lax operator (\ref{laxkdv}).
Then under the successive DBT, adjoint DBT and binary DBT
of Proposition 6, the $n$-step transformed coefficients are given by
\be
u_1^{(n)}-u_1=x(q-1)(u_0^{(n)}-u_0)
\ee
with
\bea
u_0^{(n)}&=&u_0+\pa_q\pa_{t_1}\ln W_q[\phi_1,\cdots,\phi_n]
\th(W_q[\phi_1,\cdots,\phi_n]),
\qquad (\mbox{DBT})
\label{ndb}\\
u_0^{(n)}&=&u_0+\pa_q\pa_{t_1}\ln\tilde{W}_q[\psi_1,\cdots,\psi_n]
\th^{-1}(\tilde{W}_q[\psi_1,\cdots,\psi_n])
,\qquad (\mbox{adjoint DBT})
\label{nadb}\\
u_0^{(n)}&=&u_0+\pa_q\pa_{t_1}\ln\det\Om\th(\det\Om)
,\qquad (\mbox{binary DBT})
\label{nbdb}
\eea
Eqs.(\ref{ndb})-(\ref{nbdb}) provide us a convenient way to 
construct new solutions from the old ones. Especially,
starting from the trivial solution ($u_1=u_0=0$)
we can obtain nontrivial multi-soliton solutions just by
putting the (adjoint) wave functions into the formulas
(\ref{ndb})-(\ref{nbdb}). For example, the wave functions 
$\phi_i$ ($i=1,\cdots$,n) associated with the trivial
Lax operator $L=\pa_q^2$ satisfy
\be
\pa_q^2\phi_i=p_i^2\phi_i,\qquad \pa_{t_n}\phi_i=(\pa_q^n\phi_i)
\qquad p_i\neq p_j
\ee
which give the following solutions
\be
\phi_i(x,t)=E_q(p_ix)\exp(\sum_{k=0}^{\infty}p_i^{2k+1}t_{2k+1})+
\gamma_i E_q(-p_ix)\exp(-\sum_{k=0}^{\infty}p_i^{2k+1}t_{2k+1})
\label{solution}
\ee
where $\ga_i$ are constants and  $E_q(x)$ denotes the $q$-exponential function  
which satisfies $\pa_qE_q(px)=pE_q(px)$ and has the following 
representation:
\be
E_q(x)=\exp[\sum_{k=1}^{\infty}\frac{(1-q)^k}{k(1-q^k)}x^k]
\ee
Substituting (\ref{solution}) into (\ref{ndb})-(\ref{nbdb})
 gives us $n$-solition solutions.
Note that for the soliton solutions constructed out from
the trivial one ($u_1=u_0=0$)  as described above,  
they satisfy a simple relation: $u_1^{(n)}=x(q-1)u_0^{(n)}$. 
This is just the case considered by Haine and Iliev in \cite{HI,Ib}. 
In general, it can be shown \cite{AHM} that the solutions of the $q$-KdV 
hierarchy can be expressed in terms of a single function $\tau_q$ called 
tau-function such that
\bea
u_1(x,t)&=&\pa_{t_1}\ln\frac{\th^2(\tau_q)}{\tau_q}\no\\
&=&x(q-1)\pa_q\pa_{t_1}\ln\tau_q(x,t)\th(\tau_q(x,t))
\label{tau}
\eea
Eq. (\ref{tau}) together with Proposition 7 shows that
for a given solution (or $\tau_q$) of the $q$-KdV hierarchy,
the transformation properties of $\tau_q$ can be summarized
as follows
\bea
\tau_q &\to& \tau_q^{(n)}=W_q[\phi_1,\cdots,\phi_n]\cdot\tau_q
\qquad (\mbox{DBT})\no\\ 
\tau_q &\to& \tau_q^{(n)}=\th^{-1}(\tilde{W}_q[\psi_1,\cdots,\psi_n])
\cdot\tau_q\qquad (\mbox{adjoint DBT})
\label{transform}\\ 
\tau_q &\to& \tau_q^{(n)}=\det\Om\cdot\tau_q
\qquad (\mbox{binary DBT})\no
\eea
which implies that the Wronskian-type (or binary-type) tau-function 
can be viewed as the $n$-step transformed tau-function constructed 
from the trivial solution ($\tau_q=1$).

\section{concluding remarks}
We have constructed the elementary DBTs for the $q$-KdV hierarchy, which
preserve the form of the Lax operator and are compatible with the Lax equations.
Iterated application of these elementary DBTs produces new soliton solutions 
(tau-functions ) of the $q$-KdV hierarchy out of given ones. Following the
similar treatment and using the tau-function representation 
$u_{N-1}=\pa_{t_1}\ln\th^N(\tau_q)/\tau_q$ \cite{AHM} for $N>2$ 
we can reach the same result as (\ref{transform}) except now
$\pa\tau_q/\pa_{t_{iN}}=0$.

In fact these DBTs can be applied to the $q$-deformed KP hierarchy 
without difficulty by considering the Lax operator of the form 
$L=\pa_q+\sum_{i=0}^{\infty}u_i\pa_q^{-i}$. The
$q$-KdV  is just a reduction of the $q$-KP by imposing
the condition $(L^N)_+=L^N$.
Since the ordinary KP hierarchy admits other reductions which 
are also integrable. Hence it is quite natural to ask whether 
there exist $q$-analogue of these reductions and what
are the DBTs associated with them?
We hope we can answer this question in the near future.

{\bf Acknowledgments\/}
This work is supported by the National Science Council
of the Republic of China under Grant 
Nunbers NSC 88-2811-M-194-0003(MHT) and 
NSC 88-2112-M-009-008 (JCS)

\end{document}